\documentstyle[preprint,aps]{revtex}

\begin{document}
\draft
%
%
\title{Electronic Theory for the Nonlinear Magneto-Optical Response of
Transition-Metals at Surfaces and Interfaces:
Dependence of the Kerr-Rotation on Polarization and on the Magnetic Easy Axis}
\author{W. H\"ubner and K. H. Bennemann}
\address{Institute for Theoretical Physics, Freie Universit\"at
Berlin, Arnimallee 14, D-14195 Berlin, Germany}
\date{\today}
\maketitle
\begin{abstract}
We extend our previous study of the polarization dependence of the
nonlinear optical response
to the case of magnetic surfaces and buried magnetic interfaces.
We calculate for the longitudinal and polar configuration the nonlinear
magneto-optical Kerr rotation angle. In particular, we show which tensor
elements of the susceptibilities are involved in the enhancement of the Kerr
rotation in nonlinear optics for different configurations and we demonstrate
by a detailed analysis how the direction of the magnetization and thus the
easy axis at surfaces and buried interfaces can be determined from the
polarization dependence of the nonlinear magneto-optical response,
since the nonlinear Kerr rotation is sensitive to the electromagnetic
field components instead of merely the intensities. We also prove
from the microscopic treatment of spin-orbit coupling that there is an
intrinsic phase difference of 90$^{\circ }$ between tensor elements which are
even or odd under magnetization reversal in contrast to
linear magneto-optics. Finally, we compare our results
with several experiments on Co/Cu films and on Co/Au and Fe/Cr multilayers.
We conclude that the
nonlinear magneto-optical Kerr-effect determines uniquely the
magnetic structure and in particular the magnetic easy axis in films and at
multilayer interfaces.
\end{abstract}
\noindent
{\bf I. Introduction}\\
Low-dimensional magnetic structures such as surfaces, thin magnetic films,
and multilayer sandwiches have recently become an exciting new field of
research and applications~\cite{leo}. Especially thin magnetic films and
multilayers exhibit a rich variety of properties not previously found in bulk
magnetism such as enhanced or reduced moments~\cite{weller}, oscillatory
exchange coupling through nonmagnetic spacers~\cite{grunberg,pierce,parkin},
giant magnetoresistance~\cite{fert}, and spin-polarized quantum well
states~\cite{ortega,rader}. In particular, the observation of
a perpendicular easy axis has attracted a great deal of interest, since this
phenomenon, which cannot occur at bulk surfaces, leads to enhanced stray fields
implying small magnetic domains. These small domain structures in turn may
be applied for high density magnetic recording like ``perpendicular
recording'' devices. Thus, it is of considerable importance to characterize
the easy axis of thin magnetic films and multilayer structures with buried
interfaces by destructionless remote sensing. Conventional techniques
for probing magnetic anisotropy are usually bulk probes such as ferromagnetic
resonance (FMR)~\cite{schulz} or the magneto-optic Kerr effect
(MOKE)~\cite{bader,weller1} deducing the magnetic surface signal from an
overall bulk signal background
which requires the absence of magnetism in the remainder of the sample.
In a multilayer situation, however, with several magnetic layers present,
these probes are inadequate to measure the magnetic signal from buried
interfaces~\cite{ecsnote}. At present, the nonlinear magneto-optical
Kerr-effect (NOLIMOKE)~\cite{hub89,pan,reif,reifrau} provides a unique
and destructionless tool for the probing of buried interface magnetism by
remote sensing~\cite{theo2,kirschner}.

In previous theoretical studies we calculated the nonlinear magneto-optical
Kerr {\em spectra} by an electronic theory and showed in detail that they are
fingerprints of the structural,  electronic, and magnetic properties of
surfaces, interfaces, and films~\cite{hub}. Very importantly, we proposed on
the
basis of a detailed calculation for the longitudinal configuration with
in plane magnetization and $p$ polarization of the incident light that the Kerr
rotation in nonlinear optics is generally enhanced by one order of magnitude,
since in nonlinear optics there is no suppression of the rotation
by nonmagnetic excitations in contrast to linear optics~\cite{pusto}. This
fundamental difference results from the nonlinear polarization entering
the wave equation as the source term for second harmonic generation (SHG) thus
making this equation inhomogeneous rather than just rendering the homogeneous
wave equation at the doubled frequency. Therefore, NOLIMOKE reveals the
magnetism of surfaces, interfaces, thin films and multilayers much more
drastically than linear MOKE does for bulk magnetism. Already
several experiments confirmed our theory and the sensitivity
of NOLIMOKE for surface and interface magnetism,
and in particular the drastic enhancement of the nonlinear Kerr
rotation.

Hence, in this paper, we extend our calculation of the nonlinear Kerr rotation
to arbitrary Kerr configurations with arbitrary angles of incidence, arbitrary
input polarization and arbitrary polarization of the reflected second harmonic
(SH) light: We analyze the case of in plane magnetization as well as
magnetization perpendicular to the interface, in order to show how the strength
of this enhancement of the nonlinear Kerr rotation depends on the Kerr
configuration, the polarization of the incoming fundamental and outgoing SH
light, and on the direction of the magnetization vector
and thus on the {\em symmetry} of the nonlinear magneto-optical suceptibility
tensor. We use then the results to propose a new method using NOLIMOKE for the
determination of the magnetization direction and hence the magnetic easy axis
at interfaces, and the spin configuration in multilayer sandwiches.

Note, in our previous calculation~\cite{klaus,klausdiss} of the symmetry
dependence of SHG the ferromagnetism of the transition metals
has been neglected. Thus, in this paper, we extend our previous theory
for the polarization dependence of SHG at noble and in particular
transition metal surfaces by including the
symmetry properties of NOLIMOKE imposed by magnetism and thus (i) to calculate
the polarization dependence of the nonlinear Kerr rotation, and (ii) to
determine the direction of the magnetization vector at interfaces of films and
multilayers. In view of our previous results for the polarization dependence
of SHG we expect an interesting dependence of the enhancement of the
nonlinear Kerr rotation on the polarization and the easy axis.
Note, the polarization is controlled by the matrix elements and thus it depends
sensitively on the symmetry of the wave function~\cite{klaus,note}.

The paper is organized as follows: In section II we present our theory for the
nonlinear Kerr rotation for various configurations and in section III we
discuss our results with respect to the determination of the easy axis from
the nonlinear Kerr effect and the
magnetic phase shift in nonlinear optics. Finally, we summarize our work.\\
\newpage
\noindent
{\bf II. Theory}\\
We begin our theory by the derivation of the
nonlinear Kerr angle using the wave equation
\begin{equation}
\mbox{\boldmath $\nabla $} \times \mbox{\boldmath $\nabla $} \times
{\bf E}^{(2\omega )}\;+\;\frac{\varepsilon (2\omega )}{c^{2}}
\frac{\partial^{2}}
{\partial t^{2}}{\bf E}^{(2\omega )}\;=\;-\frac{1}
{\varepsilon_{0}c^{2}}\frac{\partial^{2}}
{\partial t^{2}}{\bf P}^{(2\omega )}\;,
\label{cc}
\end{equation}
with vacuum permittivity $\varepsilon_{0}$ and the nonlinear
polarization
\begin{equation}
\frac{1}{\varepsilon_{0}}{\bf P}^{(2\omega )}\;=\;
\chi^{(2)}(2\omega ):{\bf E}^{(1)}
(\omega )\cdot{\bf E}^{(1)}(\omega )
\end{equation}
as a source term. Furthermore, the law of reflection is used and
the complex field amplitude ${\bf E}_{r,a}^{(2\omega )}$
is decomposed into left- and right-handed
circularly polarized
light ${\bf E}_{r,a}^{(\pm )(2\omega )}$ due to the magnetic
birefringence. Then, one obtains for the complex
Kerr rotation in nonlinear optics with real part $\phi _{K}^{(2)}$
and ellipticity $\varepsilon_{K}^{(2)}$
\begin{equation}
\tan \psi_{K}^{(2)}\;=\;\tan(\phi _{K}^{(2)}\;+
\;i\varepsilon _{K}^{(2)})\;=
\;i\frac{{\bf E}_{r,a}^{(+)(2\omega )}\;-\;
{\bf E}_{r,a}^{(-)(2\omega )}}
{{\bf E}_{r,a}^{(+)(2\omega )} \;+\;
{\bf E}_{r,a}^{(-)(2\omega )}}.
\label{ab}
\end{equation}
Using for the nonlinear susceptibilities a similar decomposition
\begin{equation}
\chi _{ijk}^{(2)\pm }=\chi _{ijk,0}^{(2)}\pm \chi _{ijk,1}^{(2)}
\end{equation}
with $\chi _{ijk,0}^{(2)}$ and $\chi _{ijk,1}^{(2)}$ referring to the nonlinear
tensor elements being even or odd under magnetization reversal, respectively,
one way rewrite Eq. (3) for not too large (but appreciable) nonlinear
Kerr rotations as
\begin{equation}
\phi _{K}^{(2)}\;\approx\;
Re\frac{E^{(2\omega )}_{\varphi}(s-SH)}{E^{(2\omega )}_{\varphi}(p-SH)}.
\end{equation}
Here, $E^{(2\omega )}_{\varphi}(s-SH)$ and $E^{(2\omega )}_{\varphi}(p-SH)$
denote the reflected SH field amplidudes polarized perpendicular ($s$) to or
in the optical plane ($p$), respectively, and both resulting from incident
light
of polarization angle $\varphi $.
Using electrodynamical theory these fields $E^{(2\omega )}_{\varphi}$
are expressed by the nonlinear susceptibilities, which are then
determined by electronic theory.
Note that in the subsequent analysis we will take into account {\em all}
nonvanishing nonlinear tensor elements of the longitudinal and polar
Kerr configurations.

The above formula Eq. (5) for the Kerr rotation in nonlinear optics follows
(for not too large nonlinear
Kerr rotations) also from the expression given by Koopmans and
Rasing~\cite{theo1}.
To calculate the fields
$E^{(2\omega )}_{\varphi}(s-SH)$ and $E^{(2\omega )}_{\varphi}(p-SH)$,
we extend our previous work on the symmetry properties
and polarization dependence of optical SHG to the magnetic case.\\
The reflected light at frequency $2\omega $ is given as shown by
B\"ohmer {\em et al.}~\cite{klaus,klausdiss} for (001) surfaces and
interfaces of cubic (fcc or bcc) crystals
(with $z$ being the surface normal and C$_{4v}$ symmetry group)
without magnetization~\cite{misprint} by
\begin{eqnarray}
&&E^{(2\omega)}(\Phi,\varphi)\;=
\;2i(\frac{\omega}{c})\mid E^{(\omega)}_{0}\mid^{2}
\times\nonumber\\
&&\left( \begin{array}{c}
A_{p}F_{c}\cos\Phi\\
A_{s}\sin\Phi\\
A_{p}N^{2}F_{s}\cos\Phi
\end{array}\right)
\left( \begin{array}{cccccc}
0&0&0&\mid\;0&\chi^{(2)}_{xzx}&0\\
0&0&0&\mid\;\chi^{(2)}_{xzx}&0&0\\
\chi^{(2)}_{zxx}&\chi^{(2)}_{zxx}&\chi^{(2)}_{zzz}&\mid\;0&0&0
\end{array}\right)
\left( \begin{array}{c}
f_{c}^{2}t_{p}^{2}\cos^{2}\varphi\\
t_{s}^{2}\sin^{2}\varphi\\
f_{s}^{2}t_{p}^{2}\cos^{2}\varphi\\
2f_{s}t_{p}t_{s}\cos\varphi\sin\varphi\\
2f_{c}f_{s}t_{p}^{2}\cos^{2}\varphi\\
2f_{c}t_{p}t_{s}\cos\varphi\sin\varphi
\end{array}\right).
\end{eqnarray}
Here, $\Phi $ and $\varphi $ denote the angles of polarization
of the reflected frequency doubled and of the
incident light (see Fig. 1). $f_{c,s}$ and $F_{c,s}$
are the Fresnel coefficients and $t_{s,p}$ and $T_{s,p}$ are the linear
transmission coefficients for the fundamental and frequency doubled light.
The complex indices of refraction at frequencies $\omega $
and 2$\omega $ are $n\;=\;n_{1}\;+\;ik_{1}$ and $N\;=\;n_{2}\;+\;ik_{2}$.
The Fresnel factors are
\[
f_{s}\;=\;\sin\theta/n\;,\;\;f_{c}\;=\;\sqrt{1-f_{s}^{2}}\;,
\]
and
\begin{equation}
F_{s}\;=\;\sin\Theta/N\;,\;\;F_{c}\;=\;\sqrt{1-F_{s}^{2}}\;,
\end{equation}
where $\theta $ and $\Theta $
denote the angle of incidence of the incoming light and the angle
of reflection of the reflected SHG light, respectively.
The linear transmission coefficients are given by~\cite{klaus,klausdiss,sipe}
\begin{eqnarray}
t_{p}&=&\frac{2\cos\theta}{n\cos\theta+f_{c}}\;,
\;\;t_{s}\;=\;\frac{2\cos\theta}{\cos\theta + nf_{c}}\;,\nonumber\\
\;\;T_{p}&=&\frac{2\cos\Theta}{N\cos\Theta+F_{c}}\;,
\;\;T_{s}\;=\;\frac{2\cos\Theta}{\cos\Theta + NF_{c}}
\;.
\end{eqnarray}
The corresponding amplitudes $A_{p}$ and $A_{s}$ in Eq. (6) are
\begin{equation}
A_{p}=\frac{2\pi T_{p}}{\cos\Theta}\;\;\;
\;A_{s}=\frac{2\pi T_{s}}{\cos\Theta}\;,
\end{equation}
respectively. It is interesting to note that the prefactor $\delta z$
introduced
by B\"ohmer~\cite{klausdiss} is absorbed in the tensor $\chi^{(2)}_{ijk}$.
The relationship of this factor with the skin depth is discussed
in Appendix A.

Combining now Eqs. (5) and (6) we determine the Kerr rotation
and its dependence on $\chi^{(2)}_{ijk}$ for various configurations.
First we consider the {\em longitudinal configuration} with magnetization
parallel to the interface, {\bf M} $\parallel \hat{x}$, and where the optical
plane is the {\em xz} plane.
In this configuration, the nonlinear susceptibility tensor contains
10 different nonvanishing tensor elements with 5 of them
being even under magnetization reversal
and the other 5 ones being odd~\cite{pan}. Thus, the symmetry
breaking by the magnetization induces five more tensor elements and causes
the nonmagnetic ones to become all different. The nonlinear susceptibility
tensor is given explicitly by
\begin{equation}
\left( \begin{array}{cccccc}
0&0&0&\mid\;0&\chi^{(2)}_{xzx}&\chi^{(2)}_{xxy}\\
\chi^{(2)}_{yxx}&\chi^{(2)}_{yyy}&\chi^{(2)}_{yzz}&\mid\chi^{(2)}_{yyz}&0&0\\
\chi^{(2)}_{zxx}&\chi^{(2)}_{zyy}&\chi^{(2)}_{zzz}&\mid\;\chi^{(2)}_{zyz}&0&0
\end{array}\right).
\end{equation}
In Appendix B we give the general expressions resulting from this tensor for
the reflected $p$ and $s$ polarized SH fields
generated from fields with polarization $\varphi $.
For the calculation of the nonlinear Kerr rotation it is convenient
to assume $p$ or $s$ polarization of the incident light and to detect the
rotated polarization plane in the reflected SH signal upon magnetization
reversal. Hence, one needs to know the fields for the polarization
combinations referring to the incoming and outgoing light:
$p\longrightarrow p$, $p\longrightarrow s$, $s\longrightarrow p$,
and $s\longrightarrow s$. One obtains then for the longitudinal configuration
for $p$-polarized SH light generated from $p$ polarized
incident light
\begin{equation}
E_{p}^{(2\omega )}(p-SH)\;=\;2i
\mid E_{0}^{(\omega )}\mid^{2}A_{p}t_{p}^{2}[F_{c}\chi^{(2)}_{xzx}
\cdot 2f_{c}f_{s}+
N^{2}F_{s}(\chi^{(2)}_{zxx}f_{c}^{2}+\chi^{(2)}_{zzz}f_{s}^{2})]\;,
\end{equation}
for $p$-polarized SH light generated from $s$ polarized
incident light
\begin{equation}
E_{s}^{(2\omega )}(p-SH)\;=\;2i
\mid E_{0}^{(\omega )}\mid^{2}A_{p}t_{s}^{2}N^{2}F_{s}\chi^{(2)}_{zyy}\;,
\end{equation}
for $s$-polarized SH light generated from $p$ polarized incident light
\begin{equation}
E_{p}^{(2\omega )}(s-SH)\;=\;2i
\mid E_{0}^{(\omega )}\mid^{2}A_{s}t_{p}^{2}(\chi^{(2)}_{yxx}f_{c}^{2}+
\chi^{(2)}_{yzz}f_{s}^{2})\;,
\end{equation}
and for $s$-polarized SH light generated from $s$ polarized incident light
\begin{equation}
E_{s}^{(2\omega )}(s-SH)\;=\;2i
\mid E_{0}^{(\omega )}\mid^{2}A_{s}t_{s}^{2}\chi^{(2)}_{yyy}\;.
\end{equation}
Note that {\em magnetism} occurs in all 10 nonvanishing elements of the tensor
$\chi^{(2)}_{ijk}$ and in the complex indices of refraction at the
fundamental and the SH frequency, n($\omega $) and N(2$\omega $), respectively.
The dominant nonlinear magneto-optical Kerr effect, however, results from
$\chi^{(2)}_{ijk}$, in particular from the five tensor elements
$\chi^{(2)}_{xxy}$, $\chi^{(2)}_{yxx}$, $\chi^{(2)}_{yyy}$, $\chi^{(2)}_{yzz}$,
and $\chi^{(2)}_{zyz}$, which are odd upon magnetization reversal.
Using these expressions for $E_{\varphi}^{(2\omega )}$ we find in the case of
the longitudinal (= meridional) Kerr configuration for the nonlinear Kerr
rotation of $p$-polarized incoming light\\
\begin{equation}
\phi_{K,p}^{(2)}\;=\;Re\frac{E^{(2\omega )}_{p}(s-SH)}
{E^{(2\omega )}_{p}(p-SH)}\;=\;
Re\frac{A_{s}}{A_{p}}\frac{\chi^{(2)}_{yxx}f_{c}^{2}+\chi^{(2)}_{yzz}f_{s}^{2}}
{F_{c}\chi^{(2)}_{xzx}\cdot 2f_{c}f_{s}+N^{2}F_{s}(\chi^{(2)}_{zxx}f_{c}^{2}
+\chi^{(2)}_{zzz}f_{s}^{2})}\;,
\end{equation}
and in for $s$-polarized incoming light\\
\begin{equation}
\phi_{K,s}^{(2)}\;=\;Re\frac{E^{(2\omega )}_{s}(s-SH)}
{E^{(2\omega )}_{s}(p-SH)}\;=\;
Re\frac{A_{s}}{A_{p}}\frac{\chi^{(2)}_{yyy}}{N^{2}F_{s}\chi^{(2)}_{zyy}}.
\end{equation}
It is remarkable that the
NOLIMOKE rotation measures the {\em electric field vectors} rather than
intensities. Note, in the case of the transverse (= equatorial) Kerr
configuration
({\bf M} $\parallel \hat{y}$, optical plane is the {\em xz} plane)
no Kerr rotation can be observed. Instead one measures an intensity
change upon magnetization reversal~\cite{reif,theo2} whereas the total
reflected SH intensity does not change upon magnetization reversal in the
longitudinal configuration~\cite{theo1}.

Secondly, we determine the Kerr rotation for the {\em polar configuration},
in which the magnetization is perpendicular to the surface. Now, the
optical plane is again the xz plane. However, in contrast to the
usual notion in linear optics, perpendicular incidence is not yet assumed,
since SHG behaves different for the nonlinear excitation in the interface plane
and perpendicular to it. Linear optics, however, makes no such
difference~\cite{mokenote}.
In the case of the polar configuration the nonlinear susceptibility has
8 nonvanishing tensor elements, five of which are different.
Three of these elements are even and two
($\chi^{(2)}_{xyz}$ and $\chi^{(2)}_{zxy}$) are odd in {\bf M}.
Note that the polar configuration is much more symmetric than the
longitudinal Kerr configuration thus causing more tensor elements to vanish
and to be equal. In detail, the nonlinear susceptibility
is given by
\begin{equation}
\left( \begin{array}{cccccc}
0&0&0&\mid\chi^{(2)}_{xyz}&\chi^{(2)}_{xzx}&0\\
0&0&0&\mid\chi^{(2)}_{xzx}&\chi^{(2)}_{xyz}&0\\
\chi^{(2)}_{zxx}&\chi^{(2)}_{zxx}&\chi^{(2)}_{zzz}&\mid\;0&0&\chi^{(2)}_{zxy}
\end{array}\right).
\end{equation}
The general expressions resulting from this tensor for the reflected
$p$ polarized SH field
generated from $\varphi$ input polarization
and for the $s$ polarized SH field are again given in Appendix B.
The calculation of the fields for the polarization
combinations $p\longrightarrow p$, $p\longrightarrow s$, $s\longrightarrow p$,
and $s\longrightarrow s$ gives for $p$-polarized SH light generated from
$p$ polarized incident light in the polar configuration
\begin{equation}
E_{p}^{(2\omega )}(p-SH)\;=\;2i
\mid E_{0}^{(\omega )}\mid^{2}A_{p}t_{p}^{2}
[F_{c}\chi^{(2)}_{xxz}\cdot 2f_{c}f_{s}+
N^{2}F_{s}(\chi^{(2)}_{zxx}f_{c}^{2}+\chi^{(2)}_{zzz}f_{s}^{2})]\;,
\end{equation}
for $p$-polarized SH light generated from $s$ polarized incident light
\begin{equation}
E_{s}^{(2\omega )}(p-SH)\;=\;2i
\mid E_{0}^{(\omega )}\mid^{2}A_{p}N^{2}F_{s}t_{s}^{2}\chi^{(2)}_{zxx}\;,
\end{equation}
for $s$-polarized SH light generated from $p$ polarized incident light
\begin{equation}
E_{p}^{(2\omega )}(s-SH)\;=\;2i
\mid E_{0}^{(\omega )}\mid^{2}A_{s}\chi^{(2)}_{xyz}
2f_{c}f_{s}t_{p}^{2}\;,
\end{equation}
and finally for $s$-polarized SH light generated from $s$ polarized
incident light
\begin{equation}
E_{s}^{(2\omega )}(s-SH)\;=\;0.
\end{equation}
Note that it makes no sense to consider the field
$E_{\varphi=\pi/4}^{(2\omega )}(s-SH)$
for both the polar and longitudinal Kerr configuration, since due to the
magnetization one or even both of the two quantities
$E_{p}^{(2\omega )}(s-SH)$ and
$E_{s}^{(2\omega )}(s-SH)$ are nonzero in this case
in contrast to the nonmagnetic case
where both $E_{p}^{(2\omega )}(s-SH)$ and
$E_{s}^{(2\omega )}(s-SH)$ vanish. Using the fields $E_{\varphi}^{(2\omega )}$
in the {\em polar} Kerr configuration, we obtain for the nonlinear
magneto-optical rotation in the case of $p$-polarized incident light\\
\begin{equation}
\phi_{K,p}^{(2)}\;=\;Re\frac{E^{(2\omega )}_{p}(s-SH)}
{E^{(2\omega }_{p}(p-SH)}\;=\;
Re\frac{A_{s}}{A_{p}}\frac{\chi^{(2)}_{xyz}\cdot 2f_{c}f_{s}}
{F_{c}\chi^{(2)}_{xxz}\cdot 2f_{c}f_{s}+N^{2}F_{s}(\chi^{(2)}_{zxx}f_{c}^{2}
+\chi^{(2)}_{zzz}f_{s}^{2})}
\end{equation}
and in the case of $s$-polarized incident light\\
\begin{equation}
\phi_{K,s}^{(2)}\;=\;Re\frac{E^{(2\omega )}_{s}(s-SH)}
{E^{(2\omega )}_{s}(p-SH)}\;=\;
Re\frac{A_{s}}{A_{p}}\frac {0}{N^{2}F_{s}t_{s}^{2}\chi^{(2)}_{zxx}}\;.
\end{equation}
Note, in both the longitudinal and the polar Kerr configuration and for both
$p$ and $s$ input polarization the Kerr rotation contains only odd tensor
elements in the numerator and only even tensorelements in the denominator
as generally expected and as was derived already by
Pustogowa {\em et al.}~\cite{pusto} for not too large rotation angles.
The dependence on the incident angle results here and in ref.~\cite{pusto}
exclusively from the linear optical coefficients.
Assuming $\chi^{(2)}_{yxx}\;<\;\chi^{(2)}_{yzz}$
and $\chi^{(2)}_{zxx}\;<\;\chi^{(2)}_{xzx}\;<\;\chi^{(2)}_{zzz}$, Eq. (15)
for the nonlinear Kerr rotation in the longitudinal geometry yields
in agreement with ref.~\cite{pusto}
\begin{equation}
\phi_{K,p}^{(2)}\;=\;Re\frac{A_{s}}{A_{p}N^{2}F_{s}}\frac{\chi^{(2)}_{yzz}}
{\chi^{(2)}_{zzz}}\;\approx\;\frac{1}{N\sin\theta}\frac{\chi^{(2)}_{yzz}}
{\chi^{(2)}_{zzz}} \;,
\end{equation}
if the same approximations are made. This fact is easily
seen from the $\sin^{2}\Theta$ terms in the {\em even linear} tensor elements
contributing to the denominator of $\phi_{K, p-in}^{(2)}$ and cancelling the
$\sin\Theta$ originating from the {\em odd linear} susceptibility tensor
elements.
In ref.~\cite{pusto}, however, also the magnetism in the linear optical
factors and the detailed form of the nonlinear Fresnel coefficients
belonging to the particular choice of the configuration has been included.
Thus, the effects of the
$\frac{1}{N\sin\theta}$ term are suppressed what results in a much weaker
angular dependence of $\phi_{K,p}^{(2)}$.

This completes then the determination of the polarization dependence
of the Kerr rotation in nonlinear optics. The Kerr angle is expressed
in terms of $\chi^{(2)}_{ijk}$ which may be calculated by an electronic theory.
According to Eqs. (15), (16) and (22), (23) one may determine
the easy magnetic axis from the Kerr rotation. Of course, it is
straightforward to give also expressions for the ellipticity
$\varepsilon_{K}^{(2)}\;=\;
Im \frac{E^{(2\omega )}_{\varphi}(s-SH)}
{E^{(2\omega )}_{\varphi}(p-SH)}$. In the next section we discuss in
more detail and quantitatively the Kerr rotation in nonlinear optics at
surfaces, interfaces, and in thin films. The understanding of these cases
is a prerequisite for multilayers, where additional interference
structures come into play~\cite{theo1}.
\newpage
\noindent
{\bf III. Results and Discussion}\\
We now discuss the special cases of perpendicular and grazing incidence.
For {\em perpendicular} incidence ($\theta =\Theta =0^{\circ}$)
the Fresnel factors become
\begin{equation}
f_{s}\;=\;0 \;,
f_{c}\;=\;1\;,
F_{s}\;=\;0\;,
F_{c}\;=\;1
\end{equation}
and the linear transmission coefficients simplify to
\begin{equation}
t_{p}\;=\;t_{s}\;=\;\frac{2}{1 + n}
\;,\;\;T_{p}\;=\;T_{s}\;=\;\frac{2}{1 + N}
\;.
\end{equation}
Thus, the corresponding amplitudes $A_{p}$ and $A_{s}$ in Eq. (6) are
\begin{equation}
A_{p}\;=\;A_{s}=\frac{4\pi}{1 + N}\;.
\end{equation}
Thus, the nonlinear Kerr rotation in the
{\em longitudinal}
configuration for $p$-input polarization becomes
\begin{equation}
\phi_{K,p-in,long.}^{(2)}\;=\;1\cdot Re\frac{\chi^{(2)}_{yxx}\cdot 1}
{0\cdot \chi^{(2)}_{xzx}\cdot 0\;+\;0}\;\longrightarrow\;\infty
\end{equation}
and  for $s$-input polarization
\begin{equation}
\phi_{K,s-in,long.}^{(2)}\;=\;1\cdot Re\frac{\chi^{(2)}_{yyy}\cdot 1}
{0\cdot \chi^{(2)}_{zyy}}\;\longrightarrow\;\infty.
\end{equation}
Thus, the nonlinear Kerr rotation angle becomes arbitrarily large
for perpendicular incidence. This is equally true for $p$ and for $s$ input
polarization.
Note, this divergence of the angle means according to Eq. (5) a rotation by up
to 90$^{\circ}$.
We use now Eqs. (15) and (16) to calculate the nonlinear Kerr rotation angle
for Fe in the longitudinal configuration for $p$ and $s$ polarized incident
light. Results for $\Phi_{K}^{(2)}$ are shown in Fig. 2.
These results were obtained from our microscopic theory~\cite{pusto}
for the nonlinear Kerr susceptibilities $\chi^{(2)}_{yzz}$ and
$\chi^{(2)}_{zzz}$, the spin-orbit coupling constant has been kept fixed at
50 meV, and the complex indices of refraction at 1.6 eV and 3.2 eV were taken
from Johnson and Christy~\cite{jc} ($n=2.87+i3.28$, $N=2.12+i2.50$).
For the absolute ratios and the relative phases of the complex tensor elements
we use the values $\chi^{(2)}_{xzx}=0.60\chi^{(2)}_{zzz}e^{i1.945\pi}$,
$\chi^{(2)}_{yyy}=\chi^{(2)}_{yxx}=
1.60\chi^{(2)}_{zzz}e^{i0.5\pi}\frac{\lambda_{s.o.}}
{\hbar \omega}$, and.
$\chi^{(2)}_{yzz}=\chi^{(2)}_{zzz}e^{i0.5\pi}\frac{\lambda_{s.o.}}
{\hbar \omega}$.
For the remaining quantities we use
$\chi^{(2)}_{zxx}=\chi^{(2)}_{zyy}=0.0681\chi^{(2)}_{zzz}e^{i0.505\pi}$
for the full ($\Phi_{K,p}^{(2)}$) and long-dashed ($\Phi_{K,s}^{(2)}$) curves
and
$\chi^{(2)}_{zxx}=\chi^{(2)}_{zyy}=0.0681\chi^{(2)}_{zzz}e^{i1.505\pi}$
for the short-dashed ($\Phi_{K,p}^{(2)}$) and dotted ($\Phi_{K,s}^{(2)}$)
curves.

The results of Fig. 2 clearly show the divergence of
the nonlinear Kerr rotation $\Phi_{K}^{(2)}$ for $p$ and $s$ polarized incident
light in the case of perpendicular incidence and the increased enhancement
for $s$ polarization.
This result is supported by experiments~\cite{theo1} on Fe/Cr
multilayers which find a drastic
enhancement of the nonlinear Kerr rotation in the longitudinal configuration.
Upon decreasing the angle of incidence from 75$^{\circ}$ to 15$^{\circ}$,
values of $\phi_{K, p-in}^{(2)}$ from 1$^{\circ}$ to 5$^{\circ}$
and of $\phi_{K, s-in}^{(2)}$ from 5$^{\circ}$ to 15$^{\circ}$ are observed.
Thus, in contrast to linear optics, there is no need to resort to uranium based
compounds, large external magnetic fields or low temperatures in order to
obtain arbitrarily large Kerr rotations in nonlinear optics.

The fact that the experimental values do not increase monotonously in this
range of angles of incidence is readily understood from interference effects
in multilayers and is discussed in detail by Koopmans and Rasing. As pointed
out by these authors, at surfaces and in films  $\phi_{K, p-in}^{(2)}$
and $\phi_{K, p-in}^{(2)}$ should be monotonous
functions of angle of incidence in agreement with our result.
Our result yields in addition that the absolute values and relative phases
even of relatively small tensor elements are of major importance for an
adequate description of the nonlinear Kerr rotation angle $\Phi_{K}^{(2)}$
for $p$ and $s$ polarized incident light. Sign changes (see full curve in
Fig. 2) occur even at surfaces and in thin homogeneous Fe films. Interferences
in experiment do not only result from the multilayer structure, which brings
additional symmetry requirements and new electronic states into play, but also
from the superposition of several complex tensor elements contributing to
$\Phi_{K}^{(2)}$. This might be the reason for the ``calibration problem''
quoted by Wierenga {\em et al.}~\cite{theo2}. More work has to be done on this
point both theoretically and experimentally.

Note that our theory yields the divergence of the nonlinear Kerr rotation
for perpendicular incidence in the case of $s$ and $p$ polarized incident
light. For $p$ polarization, the limit of perpendicular incidence
is interesting, since in this situation the excitation is parallel to
the magnetization {\bf M}. This is due to the odd tensor element
$\chi^{(2)}_{yxx}$, which does not vanish.
It is clear that the approximation of not too large nonlinear
Kerr rotations underlying our theory will break down in the case of strictly
perpendicular incidence. The conclusion of this result, however, will not
change. It should also be pointed out that this large enhancement of the
nonlinear Kerr rotation is due to the arrangement of odd and even tensor
elements in the rotation where the even ones in the denominator of the
formula for the nonlinear Kerr rotation cannot be excited for
perpendicular incidence due to vanishing Fresnel factors $f_{s}$ for the
fundamental polarization at frequency $\omega$.

Our theory yields that in general for finite angles of incidence
the enhancement of the nonlinear
Kerr rotation should be much more pronounced for $s$ polarized incident light,
since in $p$ polarization the denominator of the formula for
$\phi_{K, p}^{(2)}$ contains several independent contributions which tend
to decrease $\phi_{K, p}^{(2)}$. Furthermore, the excitation in
$z$ direction (in $p$ polarization) entering this denominator causes an
only moderate enhancement of $\phi_{K, p}^{(2)}$. This is in agreement
with the theoretical results by Pustogowa {\em et al.}~\cite{pusto}
and with all experiments presently available~\cite{klausdiss,theo1,kirschner2}.
Note that the microscopic theory by Pustogowa {\em et al.} treats in particular
the enhancement in the longitudinal Kerr configuration yielding a
$\phi_{K, p-in}^{(2)}$ of 2$^{\circ}$ to 4$^{\circ}$ for a Fe surface
in the optical frequency range. This prediction is experimentally confirmed
by Koopmans and Rasing. It is important to make two remarks concerning the
choice of the longitudinal configuration and $p$ input polarization
by Pustogowa {\em et al.}: (i) Only in the
longitudinal configuration linear and nonlinear Kerr rotations can be
meaningfully compared since for the polar configuration the nonlinearity
parallel to the surface depends
very much on the localization of the excited electrons and their degree of
jellium-like behavior and in the transverse configuration no rotation is
observed. Only an intensity change upon magnetization reversal will happen.
(ii) Only in the longitudinal configuration with incident $p$ polarization
the Kerr rotation angle can be defined as in linear optics
with repect to the polarization of the incident photons at frequency
$\omega$. For incident $s$ polarization as
well as for the polar Kerr configuration one has to resort to the definition
of $\phi_{K, p}^{(2)}$ as being one half of the angle by which the major
half axis of the second harmonic polarization ellipsis is rotated upon
magnetization reversal thus not referring at all to the incident beam
polarization.

Note, Pustogowa {\em et al.} calculated the nonlinear Kerr rotation including
the magnetism in the dominant tensor elements in the nonlinear as well as in
the linear susceptibilities. In particular, the $\frac{1}{\sin \Theta}$
dependence of $\phi_{K, p}^{(2)}$ in the longitudinal configuration
is also present in their theory in the linear Fresnel coefficients but strongly
suppressed by other terms in these factors. Thus, the dependence
of $\phi_{K, p}^{(2)}$ on the angle of incidence $\Theta $ comes out much
weaker in this treatment, which is particularly suited for the frequency
dependence of $\phi_{K, p-in}^{(2)}$.
In this paper, we emphasize mostly the symmetry aspects and thus the magnetism
is neglected in the linear
susceptibilities contributing to the Fresnel and transmission coefficients.
This approximation is justified by the result by Pustogowa {\em et al.}
who found the enhancement dominantly caused by the nonlinear susceptibilities.
Thus, for the symmetry considerations, all the nonlinear tensor elements are
included in our present paper.

In the {\em polar} Kerr configuration we obtain for perpendicular incidence
ill-defined formulas for both $p$ and $s$ input polarization
\begin{equation}
\phi_{K,p-in,pol.}^{(2)}\;=\phi_{K,s-in,pol.}^{(2)}\;=\;Re\cdot\frac{0}{0}.
\end{equation}
Expansion of the formulas for nearly perpendicular incidence, however,
shows that there is no $s$-polarized second harmonic signal expected
within the electric dipole approximation, but there is $p$-SH generated
from $s$ input polarization by exciting the tensor element
$\chi^{(2)}_{yxx}$, thus leading to a vanishing $\phi_{K, s}^{(2)}$.
On the other hand, for $p$-input polarization both $s$- (numerator
of $\phi_{K, p}^{(2)}$) and $p$-SH (denominator) yield become finite
(apart from accidental zeros due to interference of the several complex
tensor elements or due to a special choice of the frequency $\omega $)
thus yielding in general a finite $\phi_{K, p}^{(2)}$.

This is why NOLIMOKE is a unique tool for the determination of the easy axis
in films and at interfaces and which cannot be determined by other
tools. According to our analysis it is  necessary for that purpose to
shine light in at slightly off-perpendicular incidence. Then
{\em perpendicular} interface magnetization (see our theory for {\em polar}
Kerr configuration)
would show no nonlinear Kerr rotation for $s$ input but an appreciable Kerr
angle for $p$ input. On the other hand, the characteristic signature
of {\em in plane} magnetization (see our theory for {\em longitudinal}
Kerr configuration) is expected to exhibit a moderately enhanced nonlinear
Kerr angle for $p$ input but a large nonlinear Kerr rotation for $s$ input
polarization. Thus, in switching from $s$ to $p$ input, the nonlinear
Kerr rotation should increase for perpendicular and decrease for in plane
easy axis.

Since NOLIMOKE can also distinguish the magnetic signals coming from
different interfaces as first proposed by H\"ubner {\em et al}~\cite{hub} and
impressively experimentally detected by Wierenga {\em et
al.}~\cite{theo2,estimate}
on Co/Au sandwiches, nonlinear magneto-optics is in addition a unique
and sensitive probe for the detection and investigation of spin configurations,
in particular canted spin structures generated by oscillatory exchange coupling
in magnetic sandwich heterostructures.
For example, in the case of canted spins in neighboring layers one may choose
the light configuration such that it does not couple to the magnetization
parallel to the interface, but only to the magnetization component
perpendicular to the interface. Thus, then a canted spin configuration
may be detected. In the case of antiparallel magnetization in neighboring
thin layers 1 and 2 one has approximately for the SH yield:
\begin{equation}
I(SH)\;=\;I_{1}(M)\;+\;I_{2}(-M)\;+\;\ldots\;=\;I_{1}(M)\;+\;I_{2}(M)
\;+(I_{2}(-M)\;-\;I_{2}(M))\;.
\end{equation}
Finally we discuss the NOLIMOKE rotation for {\em grazing} incidence
$\theta\;=\;\Theta\;=\;90^{\circ}$. In this case we have $\cos\theta=0$ and
$\sin\theta=1$ and get the following Fresnel factors:
\begin{equation}
f_{s}\;=\;\frac{1}{n} \;,
f_{c}\;=\;\sqrt{1-(\frac{1}{n})^{2}}\;,
F_{s}\;=\;\frac{1}{N}\;,
F_{c}\;=\;\sqrt{1-(\frac{1}{N})^{2}}.
\end{equation}
For grazing incidence it is meaningless to consider the transmission
coefficients alone which vanish. Instead what matters for the Kerr rotation
are the amplitudes which $A_{p}$ and $A_{s}$ which become
\begin{equation}
A_{p}\;=\;\frac{4\pi}{F_{c}}\;\;,\;A_{s}=\frac{4\pi}{NF_{c}}\;.
\end{equation}
This yields the ratio
\begin{equation}
\frac{A_{s}}{A_{p}}\;=\;\frac{1}{N}.
\end{equation}
These equations together with Eqs. (15), (16) and (22), (23) show that
the use of grazing incidence does not lead to simplified formulas for the
NOLIMOKE rotation in the case of $p$ or $s$ input polarization.

Furthermore, it is an interesting observation that in NOLIMOKE there is a
relative phase of $90^{\circ }$ between the odd and even elements of the
nonlinear susceptibility tensor  $\chi^{(2)}_{ijk}$ in contrast to linear
optics. This phase has already been found in the early
theories by H\"ubner {\em et al.}~\cite{hub89} and Pan {\em et al.}~\cite{pan}
and has later been observed in the experiment by Wierenga {\em et
al.}~\cite{theo2}. They observed $\Phi =88^{\circ }$. We discuss the
microscopic origin of this relative phase in Appendix C.\\
In summary, we have shown that in SH the Kerr rotation depends sensitively
on the light polarization, of the magnitude and direction of the magnetization
and therefore on the easy axis.
\acknowledgments
The authors gratefully acknowledge stimulating discussions with K. B\"ohmer,
J. Kirschner, E. Matthias, U. Pustogowa, Th. Rasing, and R. Vollmer.
\newpage
\noindent
{\bf APPENDIX A: NONLINEAR RESPONSE DEPTH AND SKIN DEPTH}\\
Since the nonlinear optical response results only from the range of broken
electronic inversion symmetry, B\"ohmer {\em et al.} introduced in Eq. (6)
an additional artificial prefactor of $\delta z$ denoting the nonlinear
response
depth. In this appendix we briefly discuss the relationship of the prefactor
$\delta z$ introduced in ref.~\cite{klausdiss} with the skin depth and show
that this factor is already automatically contained in our microscopic
expression of the nonlinear magneto-optical susceptibility~\cite{hub}.
Thus, in the present theory this prefactor is implicitly included in
the expression given above
for $E^{(2\omega)}(\Phi,\varphi)$, since it can be combined with
the ratio $\frac{\omega}{c}$ to yield the dimensionless constant $qa$
\begin{equation}
\frac{\omega}{c}\times \delta z\;=\;qa,
\end{equation}
where $q$ is the incident photon momentum and $a$ is the lattice constant
of the material representing a typical range of broken inversion symmetry
from which the nonlinear response is generated. The factor $qa$, however,
is contained in the tensor $\chi^{(2)}_{ijk}$ and has therefore not been
written explicitly in Eq. (6) and all subsequent expressions for the
nonlinear fields. This factor describes the ratio
of linear excitation depth $\frac{1}{q}$ and nonlinear response depth $a$
{}~\cite{hub}. The linear excitation depth $\frac{1}{q}$ is connected
to the usual electrodynamical formula for the metallic skin depth
\begin{equation}
d_{skin}\;=\;\sqrt{\frac{1}{\mu_{0}\sigma \omega}}\;=\frac{1}{q}
\end{equation}
with conductivity $\sigma$ and vacuum permeability $\mu_{0}$.
This becomes for metals at frequencies below the plasma resonance $\omega_{p}$
\begin{equation}
d_{skin}\;=\;\frac{c}{\omega_{p}}.
\end{equation}
\\
{\bf APPENDIX B: REFLECTED SH FIELDS FOR ARBITRARY
POLARIZATION ANGLE OF THE INCOMING LIGHT}\\
In this appendix we give the general expressions resulting from the tensor
$\chi^{(2)}_{ijk}$ for the reflected
$p$ and $s$ polarized SH field
generated from $\varphi$ input polarization.
In the longitudinal Kerr configuration one obtains for the reflected
$p$ polarized SH field
generated from $\varphi$ input polarization
\begin{eqnarray}
E_{\varphi }^{(2\omega )}(p-SH)&=&2i
\mid E_{0}^{(\omega )}\mid^{2}A_{p}[[F_{c}\chi^{(2)}_{xzx}\cdot 2f_{c}f_{s}+
N^{2}F_{s}(\chi^{(2)}_{zxx}f_{c}^{2}+\chi^{(2)}_{zzz}f_{s}^{2})]
t_{p}^{2}\cos^{2}\varphi\nonumber\\
&&+(F_{c}\chi^{(2)}_{xxy}f_{c}+N^{2}F_{s}\chi^{(2)}_{zyz}f_{s})2t_{p}t_{s}
\cos\varphi\sin\varphi
+N^{2}F_{s}\chi^{(2)}_{zyy}t_{s}^{2}\sin^{2}\varphi ]
\end{eqnarray}
and for the reflected $s$ polarized SH field generated from $\varphi$
input polarization
\begin{eqnarray}
E_{\varphi }^{(2\omega )}(s-SH)&=&2i
\mid E_{0}^{(\omega )}\mid^{2}A_{s}
[(\chi^{(2)}_{yxx}f_{c}^{2}+\chi^{(2)}_{yzz}f_{s}^{2})t_{p}^{2}\cos^{2}\varphi
\nonumber\\
&&+\chi^{(2)}_{yyz}2f_{s}t_{p}t_{s}\cos\varphi\sin\varphi +\chi^{(2)}_{yyy}
t_{s}^{2}\sin^{2}\varphi].
\end{eqnarray}
In the polar Kerr configuration the tensor $\chi^{(2)}_{ijk}$
yields for the reflected $p$ polarized SH field
generated from $\varphi$ input polarization
\begin{eqnarray}
E_{\varphi }^{(2\omega )}(p-SH)&=&2i
\mid E_{0}^{(\omega )}\mid^{2}A_{p}[[F_{c}\chi^{(2)}_{xxz}\cdot 2f_{c}f_{s}+
N^{2}F_{s}(\chi^{(2)}_{zxx}f_{c}^{2}+\chi^{(2)}_{zzz}f_{s}^{2})]
t_{p}^{2}\cos^{2}\varphi \nonumber\\
&&+N^{2}F_{s}\chi^{(2)}_{zxx}t_{s}^{2}\sin^{2}\varphi
+(F_{c}\chi^{(2)}_{xyz}f_{s}+N^{2}F_{s}\chi^{(2)}_{zxy}f_{c})2t_{p}t_{s}
\cos\varphi\sin\varphi ]
\end{eqnarray}
and for the $s$ polarized SH field
\begin{equation}
E_{\varphi }^{(2\omega )}(s-SH)\;=\;2i
\mid E_{0}^{(\omega )}\mid^{2}A_{s}
(\chi^{(2)}_{xxz}2f_{s}t_{p}t_{s}\cos\varphi\sin\varphi+\chi^{(2)}_{xyz}
2f_{c}f_{s}t_{p}^{2}\cos^{2}\varphi).
\end{equation}
\\
{\bf APPENDIX C: MAGNETIC PHASE SHIFT IN NONLINEAR OPTICS}\\
In this appendix, we discuss the microscopic origin of the relative phase
shift of 90 $^{\circ}$ between the odd and even elements of the nonlinear
susceptibility tensor $\chi^{(2)}_{ijk}$.
First, we have to remark that this
phase does not result from the fact that the nonlinear suceptibilities
contain three matrix elements each yielding a factor of $i$ rather than two
in the linear case, since this difference occurs in the even as well as in the
odd tensor elements. Instead, the microscopic origin is due to spin-orbit
coupling which acts as a perturbation on one of the wave functions in the
matrix elements of the odd tensor elements alone.
For a plane wave basis, for example, the spin-orbit perturbation
yields the following identity~\cite{fert1,nozieres} which can be proven by
commutator algebra
\begin{equation}
\langle{\bf k^{\prime}}\mid\lambda_{s.o.}({\bf k\times s})
\nabla V\mid {\bf k}\rangle\;=\;i\lambda_{s.o.}V_{{\bf k^{\prime}}-{\bf k}}
({\bf k}\times{\bf k ^{\prime}})\cdot{\bf s},
\end{equation}
thus giving a phase factor of $i$ in the odd susceptibility tensor elements.
This argument holds in the linear as well as in the nonlinear case, but
the resulting phase of $i$ is compensated only in the linear case
by the decomposition of $\chi_{ijk}^{(1)}$ yielding another factor of $i$
\begin{equation}
\chi _{ijk}^{(1)\pm }=\chi _{ijk,0}^{(1)}\pm i\chi _{ijk,1}^{(1)}\sin\theta.
\end{equation}
This factor comes from the wave equation, which is homogeneous in linear
optics. The susceptibility results directly from the dielectric function
the square root of which is the eigenvalue of the wave equation, the
complex index of refraction. The eigenmodes are left or right handed
circularly polarized photons.
In the nonlinear case, however, the decomposition has no factor of $i$
\begin{equation}
\chi _{ijk}^{(2)\pm }=\chi _{ijk,0}^{(2)}\pm \chi _{ijk,1}^{(2)}\;,
\end{equation}
since $\chi _{ijk}^{(2))}$ is not related to the eigenvalues of the
wave equation, which in nonlinear optics is an inhomogeneous differential
equation.\\

\newpage
\begin{figure}
\caption{The polarization and geometry of the incoming $\omega $ and reflected
2$\omega $ light, respectively. $\varphi $ and $\Phi $ are the polarizations
of the incident light and the reflected frequency-doubled photons.
$\varphi =0^{\circ }$ corresponds to $p$ polarization and
$\varphi =90^{\circ }$ to $s$ polarization. $\theta $ denotes the angle of
incidence. The crystal axes
$x$ and $y$ are in the crystal-surface plane whereas $z$ is parallel to the
surface normal.}
\label{fig1}
\end{figure}
\begin{figure}
\caption{Nonlinear Kerr rotation angles for $p$ polarized incident light
$\phi^{(2)}_{K,p}$ (full and short-dashed curves) and for $s$ polarized
incident
light $\phi^{(2)}_{K,s}$ (long-dashed and dotted curves) for Fe at 770 nm as a
function of the angle of incidence $\theta $ in the longitudinal Kerr
configuration. The relative phases
between $\chi^{(2)}_{zxx}\;=\;\chi^{(2)}_{zyy}$ and $\chi^{(2)}_{zzz}$
is 0.505$\pi$ in the full and long-dashed curves and 1.505$\pi$ in the
short-dashed and dotted curves.}
\label{fig2}
\end{figure}

\begin{references}
\bibitem[11] {ecsnote} On the other hand, surface sensitive probes for
magnetism like electron capture spectroscopy cannot penetrate a sandwich
structure and therefore also provide no access to the magnetism of buried
interfaces nor are they destructionless.
\bibitem[22] {note} At the surface and at interfaces also dipole transitions
with $\Delta m =0,\pm 1$ and $\Delta l =0,\pm 2$ are allowed since
the perpendicular inversion symmetry is broken.
\bibitem[24] {misprint} Note , since there is a misprint in the $\chi^{(2)}$
of this formula in ref~\cite{klaus}, we give here the correct version.
\bibitem[26] {mokenote} Nevertheless, in linear optics the Kerr rotation
in the polar configuration is usually larger than in the longitudinal
configuration, since (i) the polar configuration always implies perpendicular
incidence and thus large trigonometric prefactors (which tend to zero for
perpendicular incidence in the longitudinal configuration~\cite{ute})
 and (ii) a somewhat enhanced surface contribution might be present
for perpendicular magnetization, which is not averaged out over the skin depth
of a metal.
\end{references}
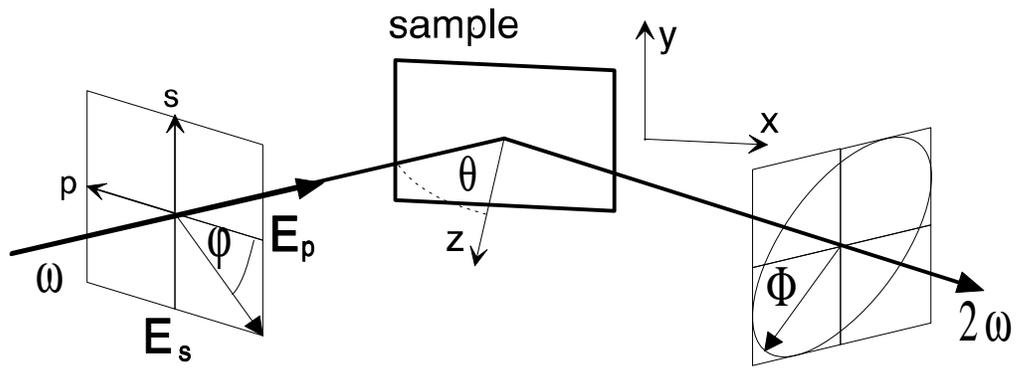
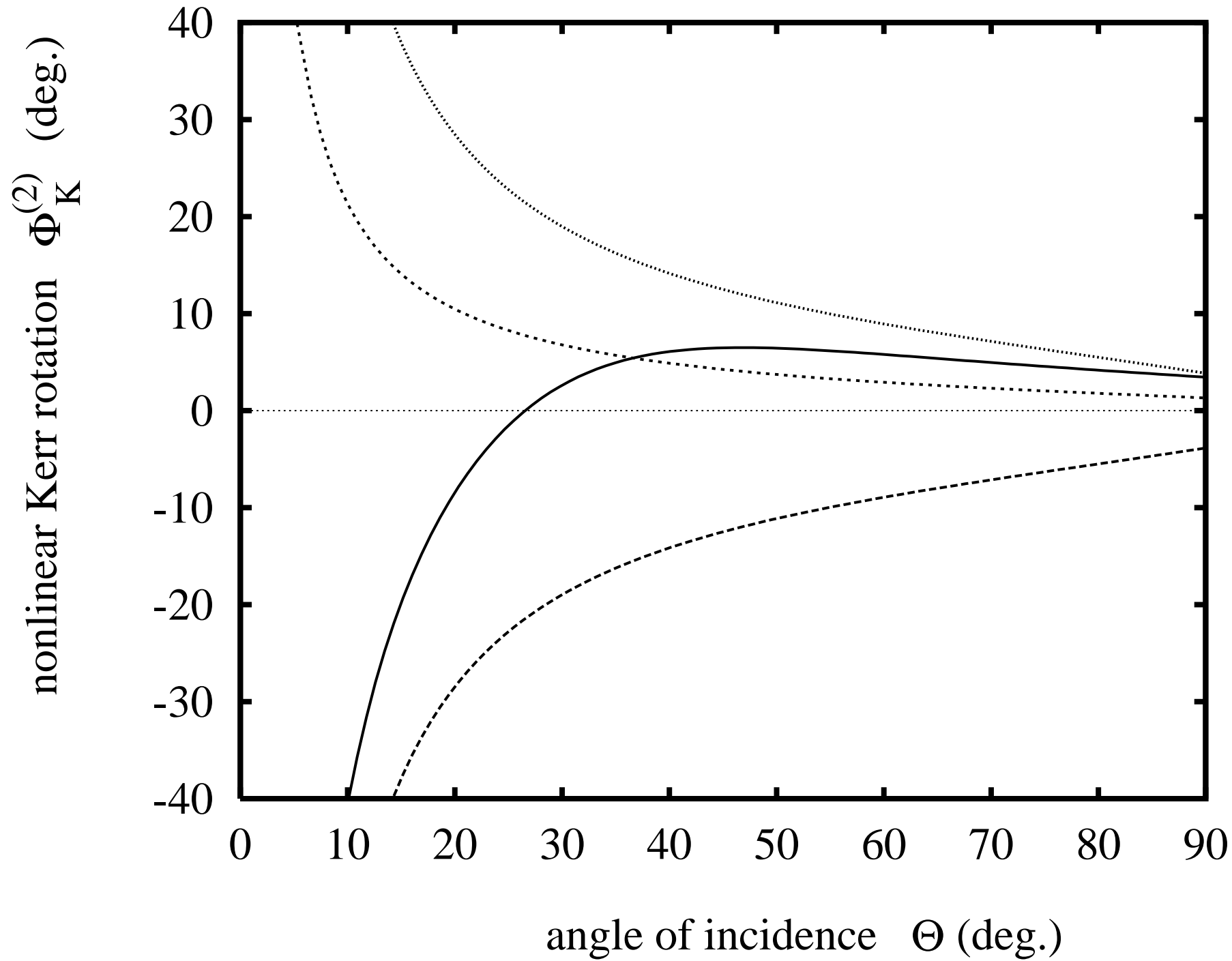
\end{document}